\documentstyle[aas2pp4,10pt]{article}

\newcommand{\etal}{{\it et~al.\ }}

\newcommand{\eg}{{\it e.g.,\ }}

\def\degr{\hbox{$^\circ$}}

\slugcomment{{\it A.J.}, in press}
\lefthead{Gladman {\it et al.} }
\righthead{Pencil-Beam Surveys for Faint Trans-Neptunian Objects}

\begin{document}

   \title{Pencil-Beam Surveys for Faint Trans-Neptunian Objects}

   \author{Brett Gladman\altaffilmark{1}}
   \affil{Canadian Institute for Theoretical Astrophysics, University 
         of Toronto, 60 St. George Street, Toronto, ON, M5S 3H8 CANADA}
   \author{JJ Kavelaars\altaffilmark{1}}
   \affil{Department of Physics and Astronomy, McMaster University,
          Hamilton, ON, L8S 4M1, CANADA}
	 	\and
   \author{Philip D.~Nicholson, Thomas J.~Loredo, Joseph A.~Burns}
   \affil{Department of Astronomy, Cornell University,
         Ithaca, NY, 14853, USA}

\altaffiltext{1}{
Visiting Astronomers: Mount Palomar and Canada-France-Hawaii Telescope.
Observations at the Palomar Observatory were made as part of a continuing 
collaborative agreement between the California Institue of Technology
and Cornell University.
The Canada-France-Hawaii Telescope is operated by
the National Research Council of Canada, le Centre National de
la Recherche Scientifique de France, and the University of Hawaii. }

   \begin{abstract}

Motivated by a desire to understand the size distribution of objects 
in the Edgeworth-Kuiper belt, an observing program has been conducted at the
Palomar 5-m and Canada-France-Hawaii 3.6-m telescopes.
We have conducted pencil-beam searches for outer solar system objects 
to a limiting magnitude of $R\sim26$.
The fields were searched using software recombinations of many short exposures
shifted at different angular rates in order to detect objects at differing 
heliocentric distances.
Five new trans-neptunian objects were detected in these searches.
Our combined data set provides an estimate of $\sim90$ trans-neptunian 
objects per square degree brighter than $\simeq 25.9$.
This estimate is a factor of 3 above the expected number of objects based on
an extrapolation of previous surveys with brighter limits, and appears
consistent with the hypothesis of a single power-law luminosity function
for the entire trans-neptunian region.
Maximum likelihood fits to all self-consistent published surveys 
with published efficiency functions predicts a cumulative sky density
$\Sigma(<R)$ obeying $\log \Sigma$ = 0.76($R$-23.4) objects per square
degree brighter than a given magnitude $R$.
               
   \end{abstract}

      \keywords{ minor planets, solar system: general, comets:general}
%

\section{Introduction and Motivation}

Since the discovery of the first so-called Edgeworth-Kuiper Belt 
object 1992 QB1 (Jewitt and Luu 1993), 
approximately 65 trans-neptunian objects 
(TNOs) have been catalogued, ranging in apparent magnitude from 
about 20 to 24.6 in R-band and in heliocentric distance from 
30 to 50 AU.
The dynamical structure and properties of objects in this region
hold signatures of the outer planet-formation process (\eg Malhotra 
1995, Morbidelli and Valsecchi 1997, Weissman and Levison 1997).
Assuming comet-like albedos of $p=0.04$, the discovered objects have 
diameters $D$ ranging from $\sim$100 to 800 km. 
Albedo uncertainties will effect this size range somewhat, with
diameters scaling as $p^{-0.5}$.

The size distribution of TNOs is of great interest.
Although originally it had been hoped that the population might
be collisionless and thus might retain the signature of the planetesimal 
formation process, recent work (Davis and Farinella 1997, Stern and 
Colwell 1997) has shown that collisional effects cannot be neglected 
over the age of the solar system.
However, knowledge of the size distribution is still important for 
understanding the link between the Kuiper Belt and both the
short-period comets (Levison and Duncan 1997) and Pluto 
(Weissman and Levison 1996). 
Dones (1997) provides an excellent review of the 
open problems in Kuiper Belt research.
The HST results of Cochran \etal (1995) provided another strong 
observational motivation by statistically detecting a very large 
population of faint trans-neptunian objects near $R=28$; 
our research was partially motivated by attempting to work at 
intermediate magnitudes.

\begin{figure}
\plotone{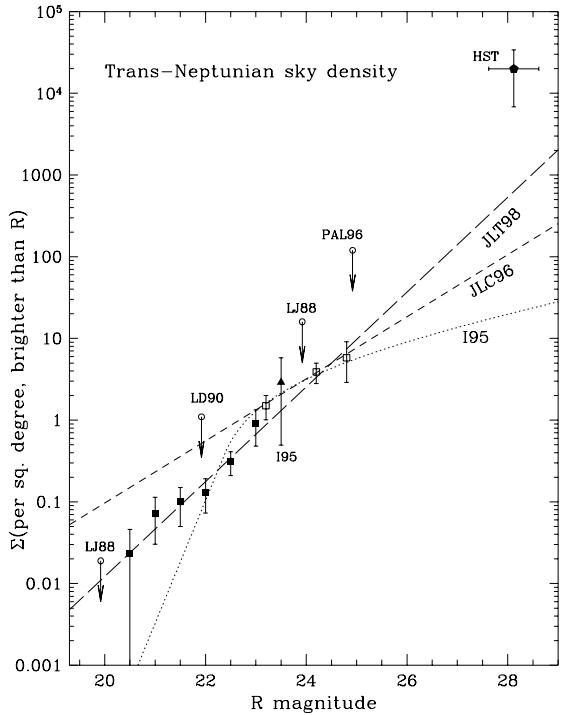} 
\caption{Previous estimates of the trans-Neptunian luminosity
        function for $R>20$, plotted as the number of objects per
        square degree near the ecliptic as a function of limiting R
        magnitude, taken from Jewitt \etal (1998) with the addition
        of the PAL 96 and HST points.
        Symbols with error bars (1-sigma) are direct observational estimates
        of the surface density of TNOs.  Solid symbols denote surveys
        where the detailed detection efficiency functions have been
        published, open symbols respresent surveys without such information.
        Open circles are 3-sigma upper limits.
The dotted line is the Irwin \etal (1995) maximum likelihood
luminosity function, fitted only to the I95 and some of the pre-1996
Jewitt \etal data.  Short-dashed line is linear fit to the Jewitt and
Luu data (open squares) and the I95 survey.
Long-dashed line is the Jewitt \etal (1998) fit to all published sky
density estimates with $R<$25 (excluding upper limits).
The HST point assumes a $V-R$ colour of 0.5 mags, and
its vertical error bar is discussed below.
         \label{fig:lfunc}
              }
 \end{figure}

Figure~\ref{fig:lfunc} shows a compilation of previous results, 
giving the cumulative surface density $\Sigma(m_R<R)$ of TNOs near the 
ecliptic brighter than a given limiting $R$ magnitude,
that is, the luminosity function of the trans-neptunian belt.
Linear relations on this plot correspond to power-law behaviour, of 
the form $\log \Sigma = \alpha (R-R_0)$, implying that the sky density 
of TNOs increases by a factor $10^\alpha$ with each additional magnitude 
of depth.
It is important to note that there is no consideration in this figure
of either the ecliptic latitude of the surveys, nor their elongation
relative to Neptune.  
The sky density of objects is expected to decline with increasing latitude 
(Jewitt \etal 1996) due to the expected concentration of the belt to 
the invariable plane of the solar system.
A peak in $\Sigma$ is expected in a magnitude-limited sample near 
90\degr\ elongation from Neptune, because TNOs trapped in
the 3:2 mean-motion resonance with Neptune have their perihelia 
concentrated near this elongation, and thus are brighter and more 
easily detected there (Malhotra 1995). 
This resonance contains $\sim40$\% of the multi-opposition objects, which 
due to the selection bias reflects an intrinsic population of only 
$\simeq$10--20\% of the TNOs inside 50~AU (Jewitt \etal 1998).
However, almost all of the catalogued objects were detected in surveys 
conducted within a few degrees of the ecliptic at elongations 
roughly $90^\circ$ from Neptune.

The goal of our program, begun in 1994, has been to find small TNOs 
rather than more objects with diameters larger than $\sim$100 km.
Instead of searching large areas of sky to limiting magnitudes of
$R\simeq 22$ -- 24, we chose to concentrate on one or two
fields for each observing run, and integrate for 4--6 hours to
reach a limiting magnitude of $R \simeq 25-26$ for each field.
In essence, the idea is that a power law increase in the sky
density of objects at fainter magnitudes will produce objects in
the field if the search is faint enough.
Thus, going deep enough will allow us to extend the range over
which the luminosity function is determined.
A previous negative result, covering 0.05 square degrees to a 
limiting magnitude of $m_R\simeq25$ was reported in Gladman
and Kavelaars (1997), and is represented as the point PAL96 in
Fig.~\ref{fig:lfunc}.  

%

The initial work of Jewitt and Luu (1995) and Irwin \etal (1995) implied
shallow slopes for the luminosity function at magnitudes of
$R\sim$23--25, with $\alpha\simeq 0.3$ (dotted line in Fig.~\ref{fig:lfunc}, 
maximum likelihood fit).
This relatively shallow slope required an upper-size
cutoff for TNOs to be consistent with the lack of observed
objects at brighter magnitudes where various surveys had failed to
find them (this information was incorporated into the Irwin \etal 
maximum likelihood fit).
Extrapolation of this shallow slope implied relatively few very faint TNOs  
($R>$28), and yet a large population of such TNOs appears to be required 
to supply the short-period comets from a Kuiper Belt source (Duncan and
Levison 1997). 
The Irwin \etal (1995) fit was in sharp disagreement with the HST result, 
which statistically detected objects near $m_R=28$, implying 
$\sim$30,000 TNOs/$\Box\degr$.
There is a continuing controversy, in the literature about the HST 
result (see Brown \etal 1997; Cochran \etal 1998), which will likely 
only be satisfactorily resolved by repeating the HST experiment.
Additional work by Jewitt and Luu (1996) doubled the number of known
objects, resulting in a steeper estimate for the 
luminosity function slope $\alpha$=0.38 (dashed line in Fig.~1, fitted 
to the hollow squares and I95 point).

More recently, Jewitt \etal (1998, JLT98 hereafter) completed a large
area survey to relatively shallow depth ($R\simeq$ 22.5), and reported a 
luminosity fuction fitted by a single power law passing through 
$\Sigma(R$$<$23.3)~=~1 object/$\Box$\degr with $\alpha$=0.58; 
{\it i.e.}, increasing by a factor of $\simeq$3.8 for each fainter 
magnitude (long-dash line in Fig.~\ref{fig:lfunc}).  
The number of detected objects at $R<22$ was far below that predicted by 
the Jewitt and Luu (1996) estimate of the luminosity function, 
implying the even steeper luminosity function.
Note that in a flux-limited survey, this steeper rate of increase implies 
that one would expect three-quarters of all detected objects to be in the 
last magnitude above the limit, {\it if} the detection limit were a strict
cut-off (in the form of a Heaviside function). 
In reality, since the discovery efficiency in a flux-limited 
sample typically falls from almost 100\% to 0 over this faintest 
magnitude above the limit, only about half of all the objects
in the final magnitude are discovered, and thus about 55--65\% of all 
discovered objects would be expected in the last magnitude above the 
limit (mildly depending on $\alpha$).
The recent survey of JLT98 (black squares in Fig.~1), corrected for 
incompleteness for the faintest magnitudes, shows exactly this behaviour; 
the previous surveys (summarized in Jewitt \etal 1996, hollow squares
in Fig.~1) did not, and we will discuss the implications of this
below.

\placetable{tbl:runs}
\begin{deluxetable}{lclccccc}
      \tablecaption{Observing Log \label{tbl:runs} }
\tablecomments{
The typical seeing $\theta$ and area $A$ for each pencil-beam survey
are given, along with the single-night integration time $T_{int}$,
the elapsed time $\Delta T$ between the first and last exposures on
that night's deep field, and instrumental zero-point.
The 50\% limit is the R magnitude at which that survey's detection
efficiency falls to 50\% (see Sec.~4); `backup' indicates that the
exposures were used to confirm candidates from the adjacent night (of
superior quality), and did not have their limits directly measured.
     }
\tablehead{
\colhead{Dates (UT)}&
\colhead{Place}  & 
\colhead{$\theta$(\arcsec)}&
\colhead{$A$($\Box$\degr)}& 
\colhead{$T_{int}$} &
\colhead{$\Delta T$} & 
\colhead{ZeroPt} & 
\colhead{50\% limit} }
\startdata
June 1995 & Palomar & 1.7 & 0.05 & 21$\times$300s & 3.03 & 25.4 & 24.8$\pm$0.2 \\
Jan.~1996 & Palomar & 2.0 & 0.05 & 38$\times$300s & 7.80 & 25.4 & 25.2$\pm$0.2 \\
April 1 1997   & CFHT    & 1.0 & 0.12 & 16$\times$480s & 4.93h & 24.6 & 24.6$\pm$0.2 \\
April 2 1997   & CFHT    & 1.2 & 0.12 & 18$\times$480s & 5.60h & 24.6 & backup April 1\\
April 3 1997   & CFHT    & 1.4 & 0.13 & 32$\times$480s & 6.66h & 24.6 & 24.6$\pm$0.2 \\
Sept.~5 1997   & Palomar & 0.9 & 0.025& 49$\times$300s & 5.78h & 25.4 & backup Sept 6\\
Sept.~6 1997   & Palomar & 0.9 & 0.025& 27$\times$480s & 5.80h & 25.4 & 25.6$\pm$0.2 \\
Sept.~7 1997   & Palomar & 0.9 & 0.025& 21$\times$480s & 3.68h & 25.4 & backup Sept 8\\
Sept.~8 1997   & Palomar & 0.9 & 0.025& 24$\times$480s & 4.38h & 25.4 & 25.6$\pm$0.2 \\
Oct.~27 1997   & Palomar & 1.4 & 0.025& 34$\times$480s & 5.80h & 25.4 & 25.2$\pm$0.2 \\
\enddata
\end{deluxetable}


Extrapolating the JLT98 luminosity function to $R=26$ predicts 
$\sim 35$ objects/$\Box\degr$, which would imply $\sim 1$ 
object per $10\arcmin \times 10\arcmin$ field 
at $R\simeq 26$ (our target magnitude).
Obviously given the steep increase of the surface density, most objects
found by such a pencil-beam search will be too faint to be followed
after the discovery observations in order to obtain an accurate orbit.
Thus, the acquisition of a larger multi-opposition orbital database of 
TNOs is {\it not} the primary goal of our work.

Under the assumptions
of a {\it constant} albedo for all objects {\it and} that 
all objects were at the same heliocentric distance $r$, then one could 
easily convert the apparent luminosity distribution into a size 
distribution.
Of course, in reality the strong $1/r^4$ dependence of reflected light 
means one must include a model of the orbital distribution
and correct for the effects of the magnitude limited sample.
Jewitt and Luu (1996) modeled this, based on surveys 
with limiting magnitudes ranging from $m_R$=23.2 to 24.8, and obtained
a size distribution in the form of a {\it differential} power law 
$n(D)\propto D^{-q}$ with index $q=3$.
The narrow magnitude range gave only a small baseline on which to 
establish such a power law.
The recent work (JLT98) to moderate limiting depth ($m_R=22.5$) but 
covering a large area of sky (52$\Box\degr$), extended the magnitude 
baseline and concluded that a steeper $D^{-4}$ size distribution was 
a better fit to the data.

One should take care to draw the distinction between the luminosity
function and the size distribution.  
The former is determined directly from observation and is relatively
free from assumptions (with awareness that $\Sigma$ will have some
dependence on ecliptic latitude and elongation relative to Neptune).
In contrast, the size distribution is dependent on many model 
parameters regarding assumptions about the orbital distribution of 
objects, their albedos, and the functional form of the 
size distribution itself.
Because of these uncertainties, we will not attempt to remodel the
size distribution to incorporate the new detections discussed below,
but rather concentrate on the luminosity function.

%

\section{Observational procedures and data reduction}

This project was carried out using observations at the
Palomar 5-meter and CFHT 3.5-meter telescopes.
Since this program was driven by the requirements of observing
very faint objects, we describe the instrumentation and data
reduction methods before discussing the results obtained during
the various observing runs of the program (Table 1).

\subsection{Instrumentation}

The 2048$\times$2048 thinned Tektronix CCD COSMIC was used at prime focus
of the 5-m Hale telescope.
The Palomar chip has high quantum efficiency (85 -- 90\%  from 550 -- 750
nm), 0.28 arcsec pixels, and a square field of view $9.7\arcmin$ on 
a side.
A Gunn {\it r} filter was used for the majority 
of the Palomar observations to minimize sky brightness; transformation
to Kron-Cousins $R$ magnitudes is straightforward.
Following the results reported in Gladman and Kavelaars (1997), two
additional Palomar observing runs occurred in the fall of 1997 (7 nights
total, 5 usable).

The University of Hawaii 8k$\times$8k prime-focus CCD array on the 
Canada-France-Hawaii 3.6-m reflector was used with a conventional KC $R$
filter.
The quantum efficiencies of this array's 8 2048$\times$4096 chips vary,
but are considerably poorer (30--40\%) than the Palomar chip, 
resulting in a search which was much 
less deep, but covered a larger field ($\simeq 30 \times 30 \arcmin$).
All CFHT results reported here are from a 3-night observing run in early 
April 1997.

Since TNOs at opposition have retrograde motions slower than 5\arcsec /hour,
integration times of 480 sec limited trailing losses while still giving
acceptable duty cycles.
For 1\arcsec \ seeing, 480 sec exposures produced a SNR of about 6 for 
objects with $R\simeq 24$ at Palomar, and with $R\simeq 22.8$ at CFHT.
Exposures at both telescopes were acquired while using an offset guide 
star.
On each photometric night we observed photometric standard fields;
NGC 7006 (Odewhan \etal 1992) from Palomar and various standard Landolt
fields (Landolt 1992) from CFHT.

\subsection{Analysis Method}

All images involved in the deep searches were preprocessed to remove
detector characteristics, including having bad columns fixed by averaging 
pixel values on either side.
Cosmic rays were not removed, for fear of removing our faint moving sources,
and because the subsequent data reduction eliminates them almost entirely.
The data analysis software consists of an 
IRAF\footnote{The Image Reduction and Analysis Facility is
distributed by the National Optical Astronomy Observatories, 
operated by the Association of Universities for Research in Astronomy, Inc.
(AURA) under cooperative agreement with the NSF.}
program which, given an angular rate and direction, recombines the images 
by shifting their pixels and then combining them.
The offset for each frame is calculated based on an assumed
drift rate and the the time delay between the start of that exposure and 
and the first frame in the sequence.
Thus, all stationary objects will elongate, trailing in the 
direction of recombination; only objects moving near the specified
angular rate will have their signal constructively add into a
single seeing disk.
Each angular rate $\dot{\theta}$ corresponds to first order to 
a different heliocentric distance.
By recombining the frames at a variety of rates, we can search for
objects from 10--100 AU.
Since there appear to be many fewer Centaurs per square degree than TNOs
(Jewitt and Luu 1996), the most likely discovery is of new TNOs 
in the range 30--50 AU.
Our 480 sec integration time $T$ resulted in trailing losses only inside
18 AU, using the empirical trailing-onset criterion (Jewitt and Luu
1996) of $\dot{\theta}$T/(FHWM) $>$ 1, in seeing with a FWHM 
of 0.9\arcsec .

One would like to remove all fixed sources from these frames before
beginning the processing. 
Experimentation with subtracting from each frame a median image created
from all the un-shifted frames met with mixed success due
to the problem of variable seeing (over 4--6 hours of integration),
causing different point-spread functions for stellar images.
Since the deep-search fields were selected to have few background
stars (a few percent of the field area), this refinement produced 
negligible improvement, and thus was not used.
Instead we settled on an algorithm which created two analysis images for each 
angular rate and direction considered. 

$\bullet$
Each frame had its mean sky value subtracted and its flux equalized
by scaling a bright (but non-saturated) star.

$\bullet$
The images were then shifted at the assumed rate and direction, creating a
`stack' of shifted images corresonding to those two parameters.

$\bullet$
A first analysis image was then created by simply summing the shifted set; 
this has the advantage of preserving all the signal but suffers because all
cosmic rays appear in the `summed image'.

\begin{figure}
\plotone{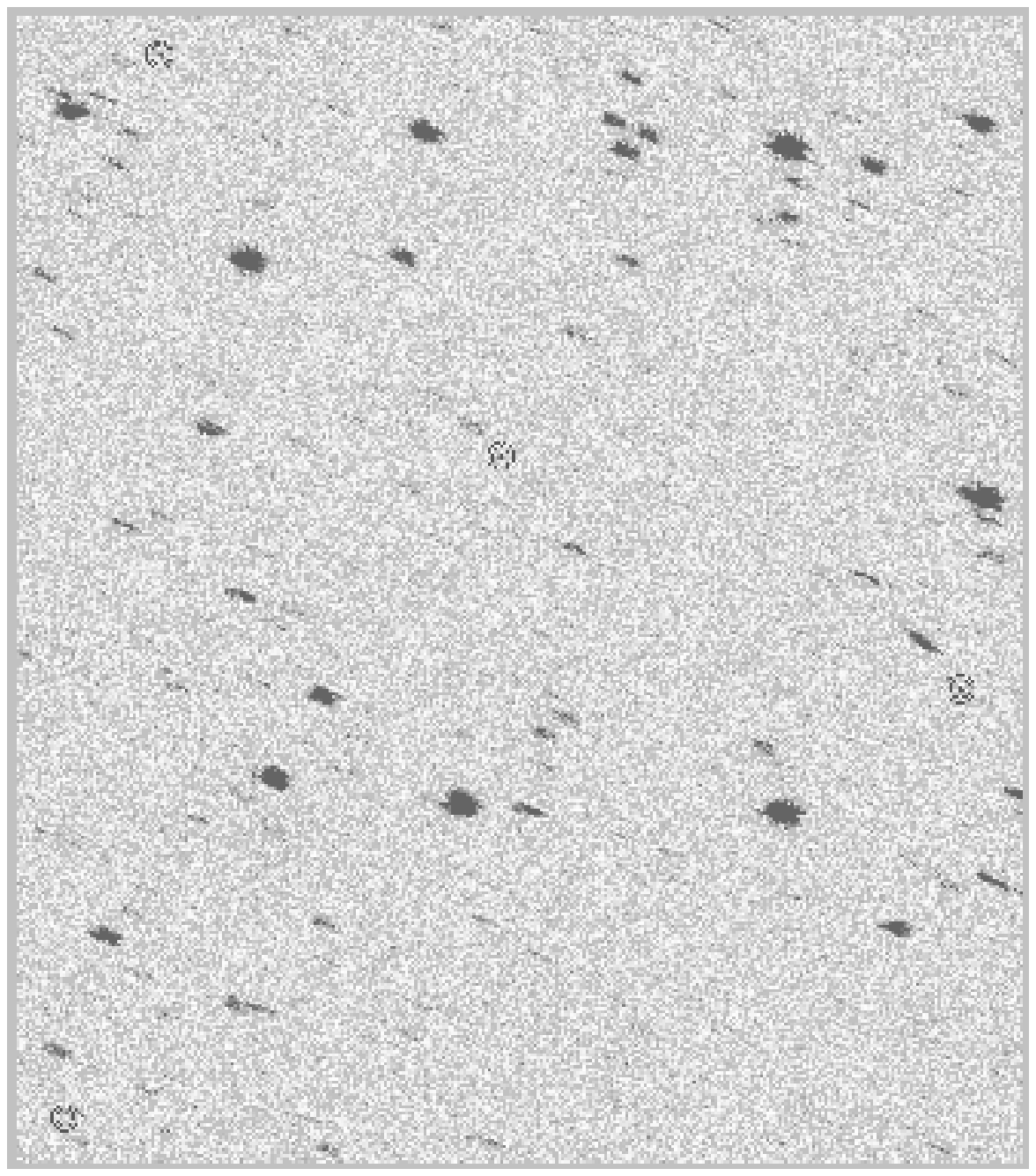} 
\caption{
A medianed image of RR20 field, from data obtained
Sept.~8/97 UT, showing 4 TNOs.
From top to bottom they are 1997 RX9, 1997 RT5, 1996 RR20, and the $m_R=25.6$
TNO pictured in Fig.~\ref{fig:wowdatfaint}.
The recombination rate is 2.9\arcsec~/hour retrograde at 23\degr\
from the equator, and
the pictured field is $\simeq$6.5\arcmin$\times$8\arcmin \ in size.
This image was constructed from twenty-five 480-sec images.
The faint, bottom-left TNO, is shown better in
Fig.~\ref{fig:wowdatfaint}.
         \label{fig:rr20median}
              }
\end{figure}

$\bullet$
The second analysis image (the `medianed image') was created by rejecting 
the highest value for 
each pixel in the shifted `stack', and then taking the median of the 
remaining pixels (\eg Fig.~\ref{fig:rr20median});
this eliminates effectively all cosmic rays, and most of the 
images of the faint stars (since non-bloomed stars contribute 
their PSF to only a small fraction of the images as they `pass by' 
in the shifting sequence). 

\noindent
It was these `medianed images' that were then {sear\-ched} for objects, 
although all candidates were examined in the `summed images' to verify 
their reality.
To search for moving objects, the analysis images were examined for
point-like objects.

What resolution is necessary in parameter space in order to detect
all objects?  
If the recombination rate (in pixels per hour) is too fast or too 
slow, the object's signal will trail into many pixels and faint 
objects will not emerge above the noise.
By experimenting with artificial objects implanted in the data, it was
determined that a grid spacing of half the seeing per hour was sufficient to 
find all objects.  
To err on the safe side, a grid spacing of one-third the seeing per hour was
used, which resulted in all objects being visible on at least two
of the medianed images (or more for brighter objects).

The orbital inclination of the objects will also produce motion
in differing directions.
However, our fields were within $10\degr$ of
opposition, implying that the motion of all objects will be heavily
dominated by their retrograde component (and means that main-belt asteroids
are far from their stationary points and cannot mimic the motion of
outer solar system objects).
Even orbital inclinations of $30\degr$ produce deviations of only
$5\degr$ in the apparent direction of TNO motion across the sky (the
dominant retrograde component being at roughly $23\degr$ to the
equator for our spring and fall observations).
The experimentally-determined grid separation required from our tests 
with implanted artificial objects was also $\simeq$5\degr.
Thus even though recombinations at solely the pure retrograde direction
should suffice, to be cautious we searched all Palomar frames at apparent 
rates corresponding to motions of 18, 23, and $28\degr$ with respect to
the equator. 
This should extend our sensitivities to all orbits with orbital 
inclinations $<$45\degr.
Since in all cases we found artificially implanted objects 
(see Sec.~4) that were moving at 18\degr\ or 28\degr\, we only
searched the CFHT results at angles of 23\degr.

We found that the most effective way to search the combined frames was to
work at constant recombination {\it direction}, and blink 4 adjacent 
{\it rates} in sequence.  
Real objects show a distinct pattern of their signal becoming stronger
and then weaker as the correct rate is approached and passed.
A best estimate of the rate and direction can then be established
by producing a much finer resolution grid in rate and direction around
the candidate object and determining which parameters maximize the 
signal.
In practice we found that the rate and direction could be determined to
similar precision by re-combining the first and second halves of the
data at the candidate rate and direction, and then measuring the 
apparent motion on these two images; real objects of course have similar
brightness on these two subsets, although the SNR is $\sqrt{2}$ smaller.
The constant brightness (to the errors inherent in the photometry) in
images created with half the data set is important, since if background 
noise were responsible for the appearance of a low-SNR object, it 
would have to be of constant amplitude along the direction of 
the trail in order to yield the same (spurious) amplitude and
brightness profile in both halves of the integration time.
All candidates had to show a profile consistent with the oversampled
PSF of the observing conditions, and exhibit the expected pattern
of changing signal as the retrograde rate was tuned.
We also examined the immediate neighbourhood of a candidate in
{\it all} the images of the set to make sure no spurious event appeared 
(such as cosmic ray strikes) which might somehow produce a signal,
even though the median process should eliminate such an eventuality. 
As a further check, each field was imaged on two adjacent nights; a
faint object at low signal to noise can thus be verified
by observing it on another night at the location where its 
measured motion should put it.
We found that we could reliably work to a SNR of about 4 with almost 
100\% detection efficiency.

%

\section{Results}

The progress reported in this paper (subsequent to the previous null
result reported in Gladman and Kavelaars 1997) comes from new data 
obtained during two observing runs at the Palomar 5-meter and one 
observing run at the CHFT 3.6-m reflector.
Since the instruments used differed, resulting in very different
limiting magnitudes and areal coverage, we will discuss the data obtained
at the two observatories separately.

\placetable{tbl:orbits}
\begin{deluxetable}{lcccccl}
      \tablecaption{Perliminary orbits for new TNOs \label{tbl:orbits}}
\tablecomments{All orbits found are outside 40 AU.
These objects have only provisional circular orbits; however, only
1997 RT5 and 1997 GA45 had sufficient observations to
rule out a 2:3 resonant orbital solution.
Radii are computed assuming geometric albedos of 4 percent.
      }
\tablehead{
\colhead{      
Designation} & 
\colhead{$m_R$} &
\colhead{$a$(AU)} & 
\colhead{$e$} & 
\colhead{$i(\degr)$} & 
\colhead{radius(km)} & 
\colhead{Orbit Reference} }
\startdata
1997 RT5  & 22.95$\pm$0.03 & 42.0 & 0.08 & 12.7 & 130 & MPEC 1997-R12, 1998-L03
\\
1997 GA45 & 23.7$\pm$0.5   & 43.9 & 0.09 & 8.3 & 80 & MPEC 1998-G10 \\
1997 RX9  & 24.0$\pm$0.1   & 42.1 & 0  & 30 & 80 & MPEC 1997-S09, MPC 30791 \\
faint TNO & 25.6$\pm$0.3   & 44.3 & 0 & $<$10 & 40 & 1 night only, 
no designation \\
1997 RL13 & 25.8$\pm$0.3   & 44.5 & 0 & $>$6  & 40 & MPEC 1998-E05 \\ 
\enddata

\end{deluxetable}


\subsection{Palomar data}

Data from two observing runs were available: 4 nights beginning 
Sept.~5 1997 UT, and from one night of a 3-night run beginning Oct.~26 
1997 UT.
Conditions during September were excellent with median seeing of 
about 0.9\arcsec\ on all 4 nights, allowing two deep fields to 
be obtained, on each of two nights under very stable conditions.
The October conditions were much poorer (seeing $\sim$ 1.3--1.7\arcsec), 
allowing only one deep field to be obtained (with a shallower magnitude 
depth).
Because of these differing depths, we shall discuss these two observing
runs separately.

The single-frame limiting magnitude (SNR=5) for the 8-minute September 
exposures was $R\simeq 24.1$.
It was considered desirable if possible to have a known TNO
in the frame as a `reference object' to be recovered, and so before going
to the telescope we examined all the fields that would contain known
TNOs near opposition, and selected two which had low densities of other
luminous objects in the APM catalogue.
The relative motion of TNOs is large enough that such fields are just
as likely to contain further new objects as are any randomly chosen field
near the ecliptic.
The previously known object is of course not counted in any estimate of
the surface density coming from detections in that field.

The first field was based on the Sept.~7.0 1997 UT position of 1996 RR20,
which was exhibiting retrograde motion near opposition at about 
2.9\arcsec/hr.
Figure~\ref{fig:rr20median} shows the median recombination created by 
shifting the Sept.~8.0 images at this rate parallel to the ecliptic.
1996 RR20 is at right center with a SNR$\sim40$ in the recombined frame.
In fact, this field contains {\it four} TNOs. 
Blinking the images at the telescope had immediately yielded a second bright
TNO ($R=23$, upper center in Fig.~\ref{fig:rr20median}), subsequently 
designated 1997 RT5, which coincidentally was seen at a simultaneous 
observing run at La Palma by A.~Fitzsimmons \etal; see Minor 
Planet E.~ectronic Circular (MPEC) 1997-R12.
An orbit based on the discovery observations (and subsequent recovery
in October) places it on a nearly circular orbit near 42 AU; the
hypothesis of a plutino-type orbit generates much larger residuals
(B.~Marsden, private communication 1997).
The next TNO was discovered immediately upon examining the first 
recombination of the frames at the RR20 rate, and subsequently 
designated 1997 RX9.
This object had $R=24.0\pm0.1$, and was clearly extended on the
summed image at the RR20 rate in such a way as to imply both a different
angular rate and direction of motion.
The best recombination yielded an angular rate of 3.1\arcsec/hr at
an angle of 27.5\degr, indicating a heliocentric distance
$\sim 40$~AU and large orbital inclination.
The discovery observations implied $(a,e,i)$ of $\simeq$(42,0,30\degr ),
which were only slightly modified to the elements of 
Table~\ref{tbl:orbits} after an October 1997 recovery by E.~Helin and 
D.~Rabinowitz at the 200-inch (see MPC 30791).
1997 RX9 is one of the faintest and most highly-inclined TNOs discovered
to date.
Both of these new objects were followed over the course of at least
4 hours on Sept.~7 and 8, yielding abundant high-precision astrometry.

\begin{figure}
\plotone{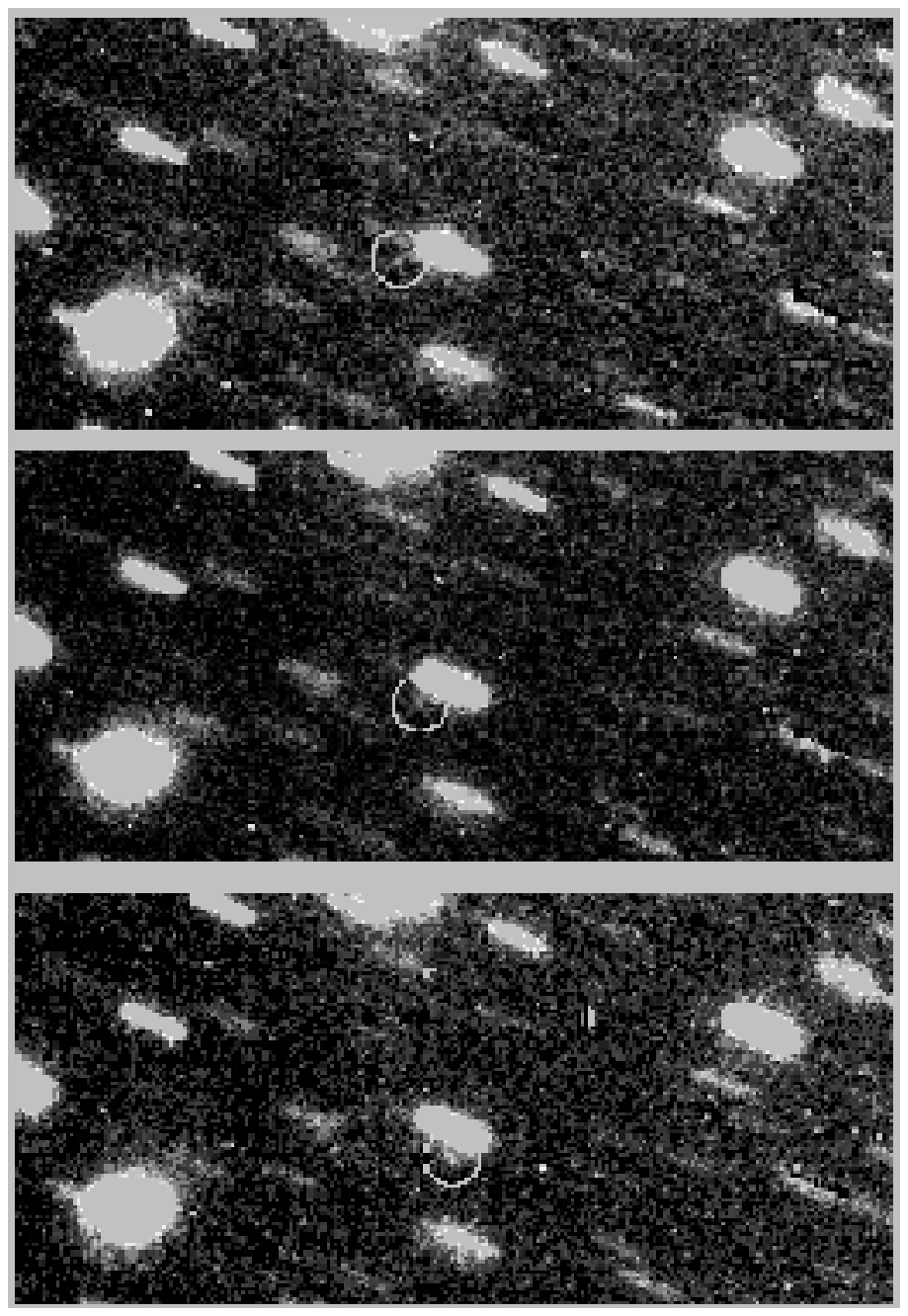} 
\caption{
The $R$=25.6 unnumbered TNO (circular orbit listed in Table~\ref{tbl:orbits}),
on recombinations summing images 1-12 (top), 7-18 (middle), and
14-25 (bottom), of the 25 images available from Sept.~8 1997 UT.
The TNO moves at a constant rate and brightness anti-parallel
to the ecliptic.
The many cosmic rays visible in the frame disappear when the
median is used (compare with Fig.~\ref{fig:rr20median}).
Recombination using all 25 frames gives
higher SNR than any of these recombinations.
         \label{fig:wowdatfaint}
}
\end{figure}

The third and final new TNO in the RR20 field was much fainter, with
$m_R\simeq 25.6$.
However, the object is easily visible in the median frames despite
the hinderance of a nearby star of moderate brightness, which it passed
during its 4-hour track on Sept.~8 UT (Fig.~\ref{fig:wowdatfaint}).
Unfortunately this object moved into the field on that night, and 
although we displaced the frames somewhat to compensate for night
to night motion, this object was approxmately 10\arcsec \ off the field
on the previous night, meaning we were unable to obtain a 2-night
arc on which to compute a preliminary orbit.
However, we have no doubt as to the reality of this object, which
appears in recombinations using only the first half, middle half, 
and second half of the frames (Fig.~\ref{fig:wowdatfaint}), moving
at a constant rate and brightness.  
The frames contain no bright pixels from cosmic ray strikes along 
the track of motion.
The PSF of the object, when combined at its correct rate of motion, 
was circular and of the stellar width, and the profile distorts
in exactly the correct pattern as the recombination rate is varied.
The retrograde motion of 2.9 \arcsec/hr indicates a heliocentric distance of
$\simeq$44 AU, with an orbital inclination of $\sim$6\degr 
(B.~Marsden, private communication) relative to the ecliptic.
Although a provisional circular orbit is reported in Table~\ref{tbl:orbits},
eccentric orbits with $a=$35--55~AU also satisfy the observations.
This object will likely never be recovered, due to its faintness and
the fact that our October nights were of insufficient quality to 
recover it in order to compute an orbit.

The field chosen for the Sept.~5 and 6 searches (observed both nights in
good conditions) was that containing the TNO 1996 TR66, this
23rd magnitude object being easily recovered.
This field yielded one new, very faint TNO ($R=25.8\pm0.3$), which 
was seen on both nights. 
The object's motion (measured via either the recombination which maximized
the signal or directly using the displacement from night to
night) implied a heliocentric distance of 44.5 AU, and the object was
subsequently designated 1997 RL13, with a provisional
circular orbit.
It should be noted that TR66, and hence this search field, was at an
ecliptic latitude of $\simeq 6\degr$; thus this is the object's minimum 
orbital inclination.
This is the faintest solar system object ever given a provisional
designation, making it unlikely that it will be recovered.
Assuming an albedo of 0.04, 1997 RL13 has a radius of only 
$\sim40$~km, making it the smallest TNO given an orbit to date.

During the four September nights we also re-obser\-ved,
the TNOs
1993 RO,
1995 QY9, 1995 QZ9, 1995 WY2,
1996 RR20,
1996 RQ20 (R-14), 1996 SZ4 (R-15), 
1996 TK66 (S-10), 1996 TR66 (S-13), 
1996 TO66, 1996 TP66, and 1996 TQ66; 
the (L-\#\#) designation after an object supplies the 1997 
Minor Planet Electronic Circular that reported the astrometry. 
The remaining observations can be found on the Minor Planet 
Circulars.
These observations providing additional information for 
improved orbital solutions.
The TNO 1997 SZ10 was discovered by D.~Jewitt 
in the same field as 1996 RQ20 on the single night of 
Sept.~24; our Sept.~7 RQ20 recovery field `pre-covered' SZ10,
allowing a 3:2 resonant orbit to be established on a 3-week baseline 
in conjunction with a recovery by C.~Hergenrother on Sept.~27
(see MPEC 1997-S16). 
We also conducted recovery attempts for the TNOs 1994 TG and 1995 YY3;
despite a limiting magnitude much deeper than their estimated
brightnesses, they were not within 5\arcmin \ of their predicted locations. 
These two TNOs are probably now lost.
This is somewhat puzzling for 1995 YY3, which was a multi-opposition
object, but our second recovery attempt in October was also 
unsuccessful.
It is possible that stellar confusion obscured the objects in each
case, but this is unlikely.

The October 1997 deep search was seriously hampered by clouds and
poor seeing,
resulting in only one deep field (0.025$\Box$\degr)
being imaged in 1.4\arcsec~median
seeing, to a limiting magnitude of $R\simeq 25.2$.
In this case we chose not to use a field with a known object present,
instead selecting a section of sky with a low background density of
objects from the Palomar Sky Survey and APM catalogue, 
at $\alpha=22^h 27^m 32^s$, $\delta = $ -6\degr
47\arcmin 30\arcsec (J2000). 
No objects were discovered in this field, which is not surprising
given the surface density estimates of $\sim$10/$\Box$\degr \ from 
previous work near this magnitude level (Jewitt \etal 1996) and
our previous upper limit (PAL96 in Fig.~1).

\subsection{CHFT data}

The CFHT data obtained with the UH8k mosaic camera resulted in 
single-frame limiting magnitudes about one magnitude shallower than 
the Palomar data.
The quantum efficiencies of two chips of the mosaic were so poor
that we did not analyze those images, meaning that we had a reduced
field of view of about 22$\times$30\arcmin , which was later
further reduced by trimming off the portions of the field not seen 
in all exposures caused by dithering.
In our April 1997 3-night observing run we imaged two fields,
one on two nights and the other on only a single night.

The first field was chosen so that one chip of the mosaic would be centered
upon an elliptical galaxy in Virgo, allowing a simultaneous study of
its globular cluster system (although half of that chip is useless for
the TNO search due to crowding in the galaxy's halo).
No new objects were found in the remaining 5.5 chips available to be searched
on the two nights of April 1 and 2 1997 UT.

The second field, imaged on only the night of April 3 UT, was chosen 
to have the object 1994 GV9 ($R$=23.1) in it. 
Although too faint to see on individual exposures, this object was 
found within a few arcseconds of its predicted position after examining 
the recombined images.
A second TNO was discovered roughly 3.5\arcmin \ northwest of GV9,
which we followed for 7 hours, and was easily visible in the recombined
frames. 
There was thus no doubt as to its reality.
The new TNO was $\sim$0.6 magnitudes fainter than GV9, and thus we 
estimate $R=23.7\pm0.5$ for the new object, although the night 
was not photometric.
This TNO would not have been given a designation except for the happy
coincidence that E.~Fletcher \etal had obtained 9 images of 1994 GV9
on April 7/8 1997 UT at La Palma, and analysis of those frames allowed a 
recovery of the new object (see MPEC 1998-G10). 
Based on the five-night arc, the orbital elements in Table~\ref{tbl:orbits} 
were derived for 1997 GA45.

Few recovery attempts were made during this observing run.
A previously un-numbered TNO was recovered, and subsequently designated 
1997 CQ29 (see MPEC 1997-J02 for details).
The mosaic's large field of view permitted the tracking of some main-belt 
asteroids over two nights, which resulted in a recovery of asteroid 2739 
Taguacipa, fortuitously moving through the field, and the discovery
and designation of a new $R\simeq 18$ asteroid 1997 GF38 (MPC 29736).

\section{Limiting magnitude determination}

\begin{figure}

\plotone{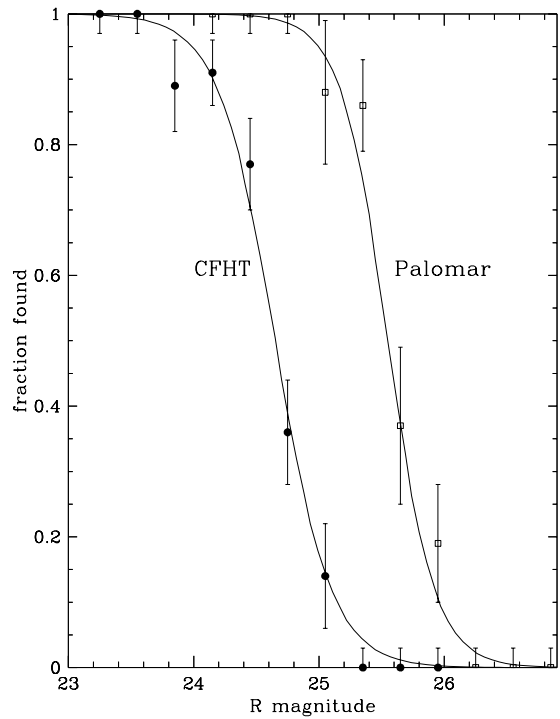} 

\figcaption[gladman.fig4.ps] {
Discovery efficiency curves for the September Palomar
data (hollow squares) and the 1994 GV9 field for the CFHT data (filled
circles).  The fraction $f$ of artificial objects found in each
magnitude bin is reported.  The fits are smooth functions based
on hyperbolic tangents which yield 50\% efficiency limits at $R$=24.6
and $R$=25.6 for the two surveys.
         \label{fig:complete}
              }
\end{figure}

After the images were aligned and the shifted images were trimmed of the
small portions which move off the frame for the given shift rate,
the Palomar frames covered an area of $\simeq$0.025 $\Box$\degr.
Therefore, 0.05 $\Box$\degr~of sky were searched in total in the
deep fields of September.
Since the single 8-minute frames had limiting 
magnitudes of $R\simeq 24$, a simple scaling indicates that the 
typically 3.5 hours of integration would be expected to yield a 
limiting magnitude of about 25.7 when recombined.
Obviously our data reduction process might not be perfectly optimal.
We established the limiting magnitude of the September data by 
writing a software algorithm that implanted a random number of
artificial objects in the data, at random places moving at random
rates in random directions (although consistent with low eccentricity
objects between 30 and 50 AU).
These frames were then searched by eye in exactly the same fashion (or
coincidentally with, in most cases) as for real objects.
The discovery efficiency followed the normal limiting-magnitude
distribution behaviour (Harris 1990), being $>95$\% until $R\simeq 25$, and 
then falling off to near zero over the next magnitude 
(Fig.~\ref{fig:complete}). 
The 50\% efficiency level is at $R=25.6\pm0.2$, which we adopt as our 
limiting magnitude.
This indicates that the shifting and recombination process has resulted
in a loss of only $\sim$0.1 magnitudes.

The detection efficiency function was determined for the October Palomar 
data in the same fashion, giving a limit of $R=25.2\pm0.2$ (since the
best October night was not photometric), over 0.025 $\Box$\degr.
An identical procedure was followed for the CFHT data, yielding a
50\% limit at $R=24.6\pm0.2$ (Fig.~\ref{fig:complete}); although 
the two fields had different numbers of exposures, the seeing 
was sufficiently worse on April 3rd that the combined image limits 
for each night are the same to within 0.05 mags.
The CFHT survey yielded 0.30$\Box$\degr\ of searched area after trimming.

Since all of our surveys have had their detection efficiency functions
evaluated (Fig.~\ref{fig:complete}), we have combined all available data 
from all our surveys to create a cumulative surface density estimate
(Table \ref{tbl:cumul}).
For convenience, we have chosen bin boundaries so that bin centers lie 
at $R$=24.6 and 25.6, where two of our surveys have 50\% completeness
points.
At each bin center in which we have detections, we have calculated
the differential surface density by taking the number of objects
and dividing by the `effective area', the latter calculated by summing the
product of the detection efficiency $f_i$ and searched area 
$A_i$ (Table 1) for each pencil-beam survey.
The resulting cumulative surface density estimates are plotted in 
Fig.~\ref{fig:lnew}.

It should be noted that the surface density estimates are based on 
the combined surveys from CFHT and Palomar, which is in principle
correct assuming no systematic errors at the level of several tenths
of a magnitude are present in the magnitudes of detected objects or 
in the determination of the efficiency functions.
If one considers {\it only} the September Palomar data set, which consisted
of 4 photometric nights in almost identical stable conditions, then our
sky density estimate becomes $\Sigma(<25.9)=120\pm60$ objects
per square degrees, being somewhat larger than that listed in Table~\ref{tbl:cumul}
since most of the detected objects are in this Palomar data set.

\placetable{tbl:cumul}
\begin{deluxetable}{cclll}
      \tablecaption{Sky densities\label{tbl:cumul}}
\tablecomments{Summary of cumulative sky densities for
the combined CFHT and Palomar deep surveys.  The uncertainty in
the `effective area' $\Sigma_i f_i A_i$ is negligible except for
the final row.  Bins are 0.5 magnitude wide, with the given centers $m_R$.
When plotted on the {\it cumulative} plot of Fig.~\ref{fig:lnew}, the faint
edge $R$ of the bin is used.  Note that these binned surface density
estimates are {\it not} used by the maximum likelihood method to compute
the luminosity function.
      }
\tablehead{      
\colhead{$m_R$}&\colhead{N}& \colhead{$\sum_i f_iA_i $} & 
\colhead{$\frac{N}{\sum_i f_iA_i}$} & \colhead{$\Sigma(<R)$} }
\startdata
23.1 & $1 \pm 100\%$ & 0.37  & $2.7 \pm 2.7 $ & $2.7\pm 2.7$ \\
23.6 & $1 \pm 100\%$ & 0.37  & $2.7 \pm 2.7 $ & $5.4 \pm 3.8$ \\
24.1 & $1 \pm 100\%$ & 0.35  & $2.9 \pm 2.9 $ & $8.3 \pm 4.8$ \\
25.6 & $2 \pm 70\%$ & $0.025 \pm 30\% $ & $ 80\pm63 $ & $90 \pm 60 $\\
\enddata

\end{deluxetable}


\section{Comparison with other surveys}

Our sky density estimates suffer the common problem of small-number
statistics and a limited magnitude range within a single survey.
In order to obtain a better estimate of the luminosity function, we
shall combine our results with those from several other published
surveys.
Care is required while doing this, since not all surveys have been
reported in the same way.

\subsection{Surface density estimates}

\begin{figure}
\plotone{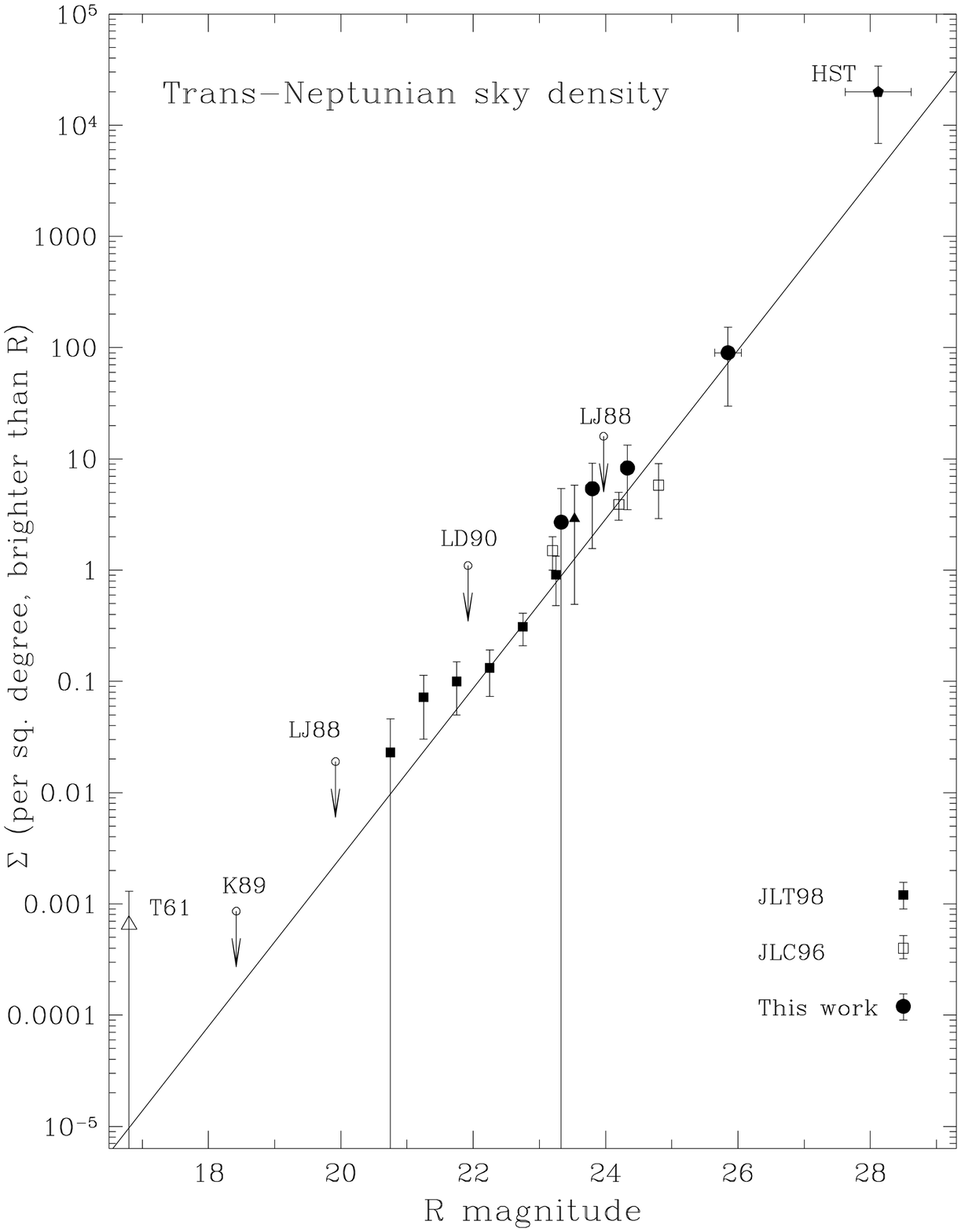} 

\figcaption[gladman.fig5.ps] {
The trans-Neptunian luminosity function, plotted as the
        cumulative number of objects per square degree near the ecliptic
        brighter than a limiting R magnitude (for 100\% completeness).
        Our new results (large circles) are listed in Table~3.
The line is the result of the maximum likelihood fit to all CCD surveys
with solid symbols and the LD90 and LJ88 CCD ($R$=24) upper limits.
The PAL96 upper limit is not shown since those pencil-beam fields have
been absorbed into our current estimates.
The JLT98 data have been shifted to the faint end of the magnitude bins.
The shown fit has $\alpha$=0.76 and $R_0$=23.4.
See text for further discussion.
         \label{fig:lnew}
              }
\end{figure}

Fig.~\ref{fig:lnew} plots estimates of $\Sigma(<R)$, along with 
our maximum likelihood fit (to be discussed below).
It is important to note that since this is a {\it cumulative} plot
of all objects brighter than a certain magnitude, the deepest edge
of a bin boundary should be used; we have thus plotted our data
(Table~\ref{tbl:cumul}) and the JLT98 data in this fashion.
All other estimates have been plotted at the stated $R$ magnitude
corresponding to 50\% completeness of the survey.  In any case, it
is not these sky density estimates that are used in the maximum
likelihood fits.

\placefigure{fig:lnew}

Our sky density estimates (Fig.~\ref{fig:lnew}) are all about 
a factor of 3 above the extrapolated JLT98 luminosity 
function.  There are several possible explanations for this (listed
below).

1. We were lucky at the 1-$\sigma$ level and the average surface
   density is lower than our determination by a factor of 3.  Note
   that we must remain systematically lucky at all of our magnitude levels.

2. Our real limiting magnitude is almost a magnitude fainter than
   our estimates.  We doubt this because we believe that the experiments
   with artificial objects clearly indicate we cannot find objects fainter
   than $R=26$ in the September data set.

3. A complex explanation could be that the surface density is not
   particularly uniform on the sky, and that the surveys all correctly
   report a {\it local} sky density.  Each survey samples a different
   depth and location relative to opposition, so this explanation is not 
   completely unreasonable.  
   Our early September opposition survey was somewhat closer to Neptune 
   than the 90\degr\ longitude separation which is usually
   searched in order to avoid the galactic plane. 
   However, the surface density of objects might be expected to {\it drop} 
   as one moves away from 90\degr\ separation from Neptune, where plutinos
   are coming to perihelion and thus increase the local surface density
   (Malhotra 1995).
   Our greater surface density works in the opposite direction from this
   expectation.  

4. The JLT98 sky density estimates, especially those for surveys 
   fainter than $R=24$, are in error.   We in fact believe this possibility
   to be the most likely, which we now discuss.
A definitive answer will require repeating the experiment, and the 
arrival of high-quality, large-format CCD mosaics will allow a 
pencil-beam survey to produce tens of detections rather than 5,
giving sufficient detections in a single field to establish
a high-quality luminosity function.

JLT98 report a luminosity function that rises by a factor of $\simeq 4$
with each additional magnitude.  As discussed above, with this
estimate of the luminosity function the number of objects in the 
final magnitude, over which the detection efficiency falls from 100\%
to zero ({\it c.f.} Fig.~4), should be roughly equal to the 
number found at all brighter magnitudes.
This is not true of the objects reported in the surveys of Jewitt \etal
(1996), and in particular for the $R=24.8$ survey, which detected
{\it none} of its 7 objects fainter than the reported $R=24.5$ 100\%
completeness point.
Thus, this survey is internally inconsistent with the derived luminosity
function at more than the 2-sigma level.
It could thus be the case that the first surveys reported in Jewitt
\etal 1996 (from $R$=23-24.8) systematically underestimate 
$\Sigma$ and should thus be viewed as lower limits or, alternately,
have stated limiting magnitudes that are too faint, as previously
suggested by Weissman and Levison 1997.
We show below that the recent JLT98 survey, which does
have 60\% of its objects at the faintest magnitudes, when analyzed
by a maximum likelihood method yields a luminosity function slope
that agrees with our results.
The HST result also suffers from the inverse internal consistency problem,
in that of order 40\% of its objects should be brighter than its 100\%
completeness limit, in contrast to the zero found.
Our estimate of the sky density estimates for the HST survey (discussed
below) incorporate this effect.

\subsection{Maximum likelihood fit to all surveys}

\def\like{{\cal L}}
\def\olike{\ell}
\def\params{{\cal P}}

We have analyzed the data from surveys available in the literature by 
adopting a simple model $\log\Sigma = \alpha(R-R_0)$,
and using Bayes's theorem to infer the model parameters.
The Bayesian approach is particularly suitable here because it allows 
us to readily combine information from disparate surveys simply by
multiplying the likelihood functions for each survey.  
The likelihood functions can be derived using the Poisson
distribution. 

\begin{figure}

\plotone{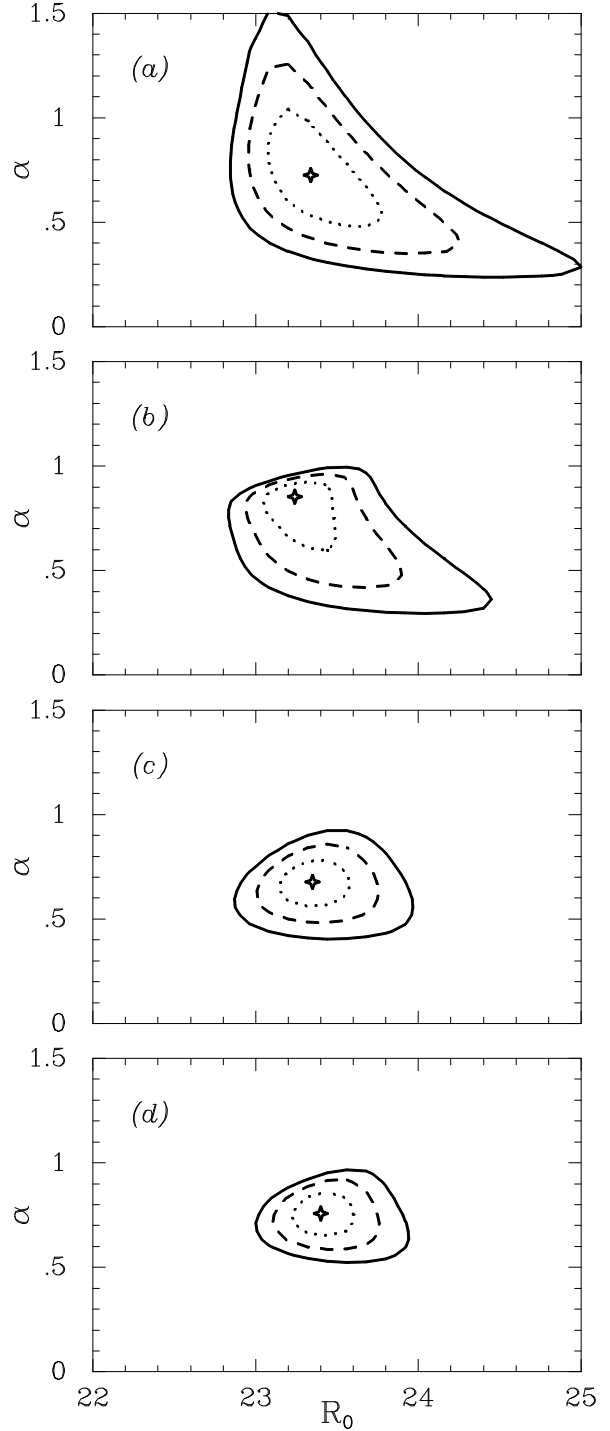} 

\figcaption[gladman.fig6.ps] {
 Best-fit parameters (cross) and credible regions
(1, 2, and 3 sigma contours) from the maximum likelihood analysis
of several CCD surveys with published efficiency functions.
{\bf a} The JLT98 survey, {\bf b} JLT98, I95, and HST, {\bf c} JLT98, I95,
with our surveys, {\bf d} all of the above with the LD90 and
JL88 upper limits.
The best fit parameters for {\bf d} are $\alpha$=0.76 and $R_0$=23.4.
         \label{fig:conf}
              }
\end{figure}

\placefigure{fig:conf}

Most surveys report a detection efficiency (or detection limit)
and the magnitudes of a finite number (possibly zero) of
detected objects.  For such surveys, the likelihood function for the
parameters $\params$ takes a form similar to that derived
by Irwin \etal (1995).  It can be written as
$$
\like_k (\params) =
\exp\!\left[ -\!\!\int\!dR\:\Omega_k\:\eta_k(R)
               \sigma(R)\right] \;
 \! \prod_i\!\int\!dR  \olike_{k i}(R) \sigma(R)
$$
where $k$ is an index denoting the survey, 
$\Omega_k (R)$ is the solid angle of survey $k$, $\eta_k (R)$
is the detection efficiency for a TNO of magnitude $R$ in that survey,
$\sigma(R)$ is the TNO surface density per unit $R$ (i.e., the
{\it differential} distribution, which depends
on the parameters $\params$), and the object likelihood function
$\olike_{k i}(R)$ describes the uncertainty for the magnitude
of object $i$ in survey $k$.  For $\eta_k (R)$, we use
fits of smooth functions to reported detection efficiencies when
the latter are reported, of the form
\begin{equation}
\eta (R)=\frac{1}{2} \left[ 1 - \tanh \left( \frac{R - R_{50}}{W} \right) 
\right] ;
\end{equation}
that is, the efficiency $\eta$ falls to 50\% at $R=R_{50}$, over a
characteristic half-width $W$.
For the upper limits we
have taken the efficiency to fall from unity to zero across a range 
of 1 magnitude centered on the quoted 50\% detection limit of the 
survey. 
For the object likelihood functions we use Gaussian functions
with means equal to the best-fit values of $R$ and standard deviations
equal to the stated errors, or the root mean squares of the statistical
and systematic errors when these are provided separately.  
This likelihood function improves on that of Irwin \etal (1995) in 
its incorporation of uncertainties and its use of the full detection 
efficiency function (Irwin \etal (1995) implicitly
took $\eta_k (R)$ to be Heaviside functions).  Of these improvements,
the latter is the most important.  A full derivation of this type
of likelihood function appears in Loredo and Wasserman (1995).

The HST survey (Cochran et al.\ 1995) did not report the magnitudes of 
individual detected objects; instead, the data
consist of counts of detected objects in each of 20 magnitude bins,
spanning a V magnitude range from  27.8 to 28.8.
We assume a $V-R$ color of 0.5; a $\pm0.5$
magnitude error bar on the detection limit is given to account for
this unknown color conversion.
The 94 detected objects are individually of low SNR and 
thus to estimate the false detection rate, 
Cochran \etal (1995) analyzed the HST images by using retrograde 
candidate orbits, and object counts (presumably of false 
objects) were reported for each bin.
A total of 65 retrograde objects were found.
We built a likelihood function for these data by modelling
the prograde and retrograde counts with Poisson counting distributions,
the former with a mean equal to the sum of the model prediction
and a background rate, and the latter with a mean equal to the
background rate.  
The background rates for each bin are unknown, but the retrograde
data provide us with estimates of them (with significant uncertainty).
The likelihood function for each bin in such an ``on-source/off-source'' 
dataset is derived in Loredo (1992).
In our analysis, we pooled the HST data into two bins (from 27.8 to 28.3,
and 28.3 to 28.8).  This was necessary because the reported
counts are so low in each bin that, taken independently, they
provide little constraint on the signal rate (i.e., the
background and signal-plus-background rates are so poorly determined in
the narrow bins that nearly every bin is consistent with zero signal
when viewed independently of other bins).  But the background
rate is presumably similar in adjacent bins, and pooling adjacent
data is a simple way to account for this.  

For the differential magnitude distribution, we ad\-op\-ted the
standard exponential model,
$$
\sigma(R) = \ln(10) \alpha 10^{\alpha(R - R_0)} ;
$$
that is, the cumulative distribution $\Sigma(F)$ obeys \linebreak
$\log\Sigma\nolinebreak=\nolinebreak\alpha(R-R_0)$, so that $R_0$ is the magnitude
where $\Sigma = 1$ TNO per square degree, and $\alpha$ is the 
slope of the distribution when plotted with log-linear axes.  
Multiplying the likelihood
functions of the various surveys produces a joint likelihood
function for $\alpha$ and $R_0$.  We adopted uniform priors,
so the posterior distribution for $\alpha$ and $R_0$ is just
the joint likelihood function, normalized, and the most probable
parameter values are simply the maximum likelihood values.
For our calculations, we normalized over the range $[0.05,2.0]$
for $\alpha$ and $[19,25]$ for $R_0$ and found credible regions
that enclose 68.3\%, 95.4\%, and 99.7\% of the total probability.
These values were chosen because of their familiarity from 
the Gaussian distribution, but the posteriors are not at all
Gaussian and the likelihood values bounding the regions have to be found
numerically; Irwin \etal (1995) presumed Gaussian statistics in finding their
parameter regions.

The JLT98 survey calculated a best-luminosity function via a least-squares
fit to the cumulatively-binned surface density estimates.
As men\-tion\-ed above, because this is a cumulative distribution this 
should be considering the estimates to be at the faint end of the 
magnitude bins.
More severely, using least-squares is incorrect for these data since (1) the
errors are Poisson, not Gaussian, and (2) the errors in the points are
highly correlated due to the cumulative distribution ({\it i.e.}, the
error for each fainter point contains the errors of all brighter ones).
We have thus re-analysed the JLT98 results using a maximum likelihood
method based on their detailed, published efficiency function for
the survey.  
We do not use the older Jewitt \etal data; the lack of objects at 
the faint end of those surveys is likely what forced the very 
shallow slope of the Irwin \etal (1995) maximum likelihood fits, 
a slope unsupported by the JLT98 data.

A cost of using a formalism that allows combination of information
from disparate surveys is that there is no simple graphical
illustration of the fitting process that precisely displays the role
of each survey in the fit.  
In Fig.~\ref{fig:lnew} we follow the common practice of plotting 
separate estimates of the cumulative TNO surface density from each 
survey, along with the cumulative density distribution corresponding 
to the best-fit model.
Such a plot must be interpreted with caution because it is {\it not possible
to construct model-independent estimates} from the data due to the 
magnitude-dependent detection efficiencies of each survey.

Particularly troublesome is representation of the HST data on such
a plot because the presence of many false detections complicates
the estimation of the surface density and its uncertainty.
For the purpose of plotting an estimate of the surface density of
objects on our figures, we fixed $\alpha$ at our best-fit value 
($\alpha = 0.76$) and rewrote the $\sigma(R)$ model, replacing the 
$R_0$ parameter with $\Sigma_{lim}$, the cumulative TNO surface 
density for a limiting magnitude of $R=28.1$.  
We then calculated the likelihood function for $\Sigma_{lim}$, and 
plotted a point with the maximum likelihood value of $\Sigma_{lim}$ 
at $R=28.1$.
The 2-bin likelihood function peaks at 20,000 per square 
degree.
The endpoints of the vertical error bar indicate where the
likelihood falls to 1/$\sqrt{e}$ its maximum value (the range spanned by
$\pm 1\sigma$ for a Gaussian likelihood).
The location of this point is not very sensitive to $\alpha$.

Fig.~\ref{fig:conf}a shows the credible regions for only the
JLT98 survey; note that very large luminosity function
slopes $\alpha$ are permitted by this data set.
We find a steeper slope ($\alpha$=0.73) than the value of 
$\alpha$=$.56\pm0.15$ given by JLT98, although that value is within 
the 1-sigma confidence level of the maximum likelihood analysis.
Both give $R_0\simeq$23.3.
Including the HST and Irwin \etal (1995) survey (Fig.~\ref{fig:conf}b) 
restricts this somewhat, especially by the HST result (ironically)
eliminating steep slopes.
Fig.~\ref{fig:conf}c shows how the best fit parameters change when 
our Palomar and CFHT survey results are included and the HST data 
are not.
Finally, Fig.~\ref{fig:conf}d gives a combined fit that also includes
all the previous surveys and the LD90 and LJ88 upper limits.
The best fit parameter values and uncertainties are 
$\alpha = 0.76^{+.10}_{-.11}$ and $R_0 = 23.40^{+.20}_{-.18}$, where 
the errors indicate the range spanned by the joint 68.3\% credible 
region in Fig.~6(d).
This implies that the cumulative sky density of TNOs increases by a factor
of $10^\alpha \simeq 6$ per magnitude.
This steeper luminosity function (Fig.~\ref{fig:lnew})
predicts more TNOs at faint magnitudes than
extrapolation of the previous I95, Jewitt \etal 1996, or JLT98 
luminosity functions, and seems to nicely bring into accordance
almost all previous surveys; this includes the formerly problematic
bright photographic surveys and the HST result.
{\it It should be noted} that the inclusion or removal of the 
HST result has very little influence on the location of the 
credible regions for the final fit (compare panel c \& d).
Clearly all of the results of these maximum likelihood fits overlap at
the 1-sigma level, but the combined data set provides a much more 
well-defined best-fit region.
There is still some uncertainty in the slope, especially interesting
because the number of faint TNOs ($R>30$) is a very strong function
of $\alpha$.
Note that because the maximum likelihood method takes into account
the 3 surveys providing upper limits, the best-fit luminosity function
`appears' somewhat low in Fig.~\ref{fig:lnew} if one looks at only
the positive detections; there are no lower limits to balance out the 
null results, and thus the sky density is pushed to lower values.

We have included on Fig.~\ref{fig:lnew} three photographic surveys
with limits $R\le20$ (Tombaugh 1961, Kowal 1989, and Luu and Jewitt 1988;
as reported in Irwin \etal 1995 and JLT98).
Although JLT98 question the validity of these surveys
(as being difficult to quantify), our best fit luminosity function
makes the non-detections by Kowal, and by Luu and Jewitt (1988), much 
less problematic than previous single power-law fits to the luminosity
function.  
Plotting Tombaugh's Pluto detection on this figure may be questionable,
since Pluto's albedo, and hence magnitude, is probably enhanced by 
its active atmosphere, meaning it may not have been detected in his 
survey if it had a dark surface.
Nevertheless, the single detection is of course formally consistent
with our luminosity function.
The question of whether there exists a maximum-size cutoff in the
size distribution (JLT98) is not directly addressed by our new results.

\section{Inside and Outside the Belt}

\subsection{Centaurs}

We also searched our two September Palomar deep fields for angular 
rates of up to 9 \arcsec/hr, corresponding to nearly circular orbits 
at about 13 AU.  This data set is free from trailing losses outside
of 18 AU, and the trailing loss mounts to 0.3 mags at 13 AU.
No new Centaurs were found in 0.05$\Box$\degr\ to magnitude R=25.6.  
Given the Jewitt \etal (1996) estimate of
$\sim$0.5 Centaurs/$\Box$\degr\ brighter than $R$=24.2, this null
result is not surprising.
Even if the sky density increases by a factor of 6 per magnitude,
at R=25.6 we expect only $\sim$6 per square degree, meaning our
0.05$\Box$\degr\ survey had only a 25\% chance of finding one.
Because of this null result, we did not search the larger but
shallower ($R\simeq 24.6$) CFHT data set for Centaurs, since it
involves looking at a large number of chips at a large number of rates,
and had only a small chance of finding any objects.

\subsection{The Belt Outside 50 AU}

We also searched the September Palomar data set for rates down to 
1.4\arcsec /hr, corresponding to heliocentric distances of nearly 100~AU.
No TNOs were found at rates lower than 2.6 \arcsec /hr, meaning we
did not observe objects at heliocentric distances greater than 50 AU,
where other surveys have also as yet failed to find any objects.
While the existence of Pluto and 1996 TL66, which journey
outside of 50 AU during their orbits, clearly implies that there are objects
in this region, we as yet have no direct observational evidence for
a `dynamically cold disk' in this region; that is, no objects on
nearly circular orbits have been found outside of 50 AU.
Is this a surprise?
Dones (1997) has discussed this issue.
 
Imagine looking in a square ecliptic field of linear size $\xi$ radians,
and corresponding linear dimension $\xi r$, where $r$ is the
heliocentric distance.
Let us assume a single power-law {\it cumulative} size distribution 
(independent of heliocentric distance) of the form 
N(diam$>D$) $\propto D^{-Q}$, and a volume number density proportional
to $r^{-\beta}$.
Note that $Q$=$q-1$, where $q$ is the differential power law.
If the surface mass density of the primordial nebula dropped as $r^{-2}$
then we expect $\beta \sim$ 2--3, consistent with constraints
derived from Monte Carlo modelling of the known TNO distribution
(JLT98).
Assume that the Kuiper belt proper ends at some inner edge $r_{min}$;
at this distance there is some minimum diameter object which can be
seen (we will be assuming constant albedos).
As we move to shells of greater heliocentric distance, the flux
from a particle of the same size drops as 1/$r^4$, and so the minimum
visible size increases as $r^2$, and thus the number of visible objects
drops as $r^{-2Q}$ due to the size variation.
The number of objects visible in the shell therefore obeys
\begin{equation}
dN \propto \xi^2 r^2 dr \ r^{-\beta} r^{-2Q}
\end{equation}
and the cumulative surface density in this field from heliocentric distance
$r_1$ to $r_2$ is
\begin{equation}
\Sigma(r_1,r_2) = \int_{r_1}^{r_2} dN/\xi^2 .
\end{equation}
We can thus derive that the fraction of objects that should be
further out than some distance $r_*$ is
\begin{equation}
\frac{\Sigma(r_*,\infty)}{\Sigma(r_{min},r_*)} = 
         \left[ \left( \frac{r_{min}}{r_*} \right)^\gamma - 1 \right]^{-1} ,
\end{equation}
where $\gamma = 5 - 2q - \beta$.

If we assume that there is no maximum diameter for TNOs, and that
the inner edge is at 30 AU, then the fraction of the objects that
should be outside $r_*$=50~AU depends heavily on the size index $q$.
Irwin \etal (1995) show that $\alpha = (q-1)/5$ if the radial distribution
is smooth; we will take $\beta=2$ although $\beta$=3 is more appropriate
for a primordial disk with constant inclination.
Dones (1997) used the shallower size distribution $q=3$, which predicts
that more than 1 quarter of all TNOs should have been discovered outside
50 AU.
For JLT98's result of $q=4$, one finds that 8\% of TNOs should be
outside 50 AU; for a somewhat steeper $q=4.8$ (from our best-fit 
$\alpha$) this drops to 4\%.
Thus, the lack of detections outside 50 AU in our pencil-beam surveys 
is not a surprise.
For the entire ensemble of $\sim 65$ TNOs, one should expect 
several or $\sim1$ such object(s), depending on the size distributions,
and on the complications introduced by a more realistic model.
For example, including a maximum diameter (meaning that large
objects do not exist to be seen) or including a `plutino'
component trapped in resonance with Neptune will both drop the
expected fraction of objects outside of 50 AU.
We also do not know if all previous surveys were uniformly
sensitive to objects moving as slowly as 2\arcsec/hour or less.
We conclude that as yet there is not a convincing problem,
and that a doubling or tripling of the TNO population will be
needed before one should begin to worry about the lack of distant
objects.

\section{Discussion}

We discovered 5 TNOs in our combined CFHT and Palomar pencil-beam
surveys.
We used a maximum likelihood analysis to combine our results with
4 other published TNO surveys.
Including our deep pencil-beam work, we conclude that the
luminosity function of TNOs is steeper than previous
estimates, with the number of TNOs increasing by a factor of 
$10^\alpha \sim 6$ per magnitude.
This rapid increase implies that a deep survey $R\sim 26$ with
a sensitive, large field of view CCD mosaic (say 0.25 square
degrees), should discover tens of TNOs in a single field;
we have been allocated observing time on the CFHT to attempt 
this project.
Our best-fit luminosity function apparently brings into accordance
almost all published faint-object surveys.
Our results neither directly confirm nor deny the validity of the HST
detections;  our best estimate of the luminosity function, if extrapolated 
to $R\simeq$28, predicts a sky density of $\sim$4,000/$\Box$\degr/, 
about 1.5-sigma below the HST estimate.

Although our original intent was to work to $R\sim$26 regardless of
the sky density, it is interesting to note that our pencil-beam 
method will actually find {\it more} TNOs for a fixed telescope 
time than the `classical' method of looking for moving objects in 
3 exposures separated by $\sim$1 hour, due to the steep luminosity 
function. 
In a background limited environment, 6 exposures are required to
work one magnitude fainter than a single exposure.
A classical search acquires one exposure of 6 different fields;
however, it must repeat the 6 fields 3 times. 
Thus, after this has been completed a deep search that concentrated
on a single field has 18 times the flux and thus goes 1.6 magnitudes 
deeper than any one of the single images from the classical method.
Thus, the ratio, $N_d/N_c$, of the number of objects discovered by the 
deep survey to the classical survey is 
\begin{equation}
\frac{N_d}{N_c} = \frac{1}{6} 10^{1.6 \alpha}
\end{equation}
where the factor of 6 appears due to the greater areal coverage of
the classical method.
The methods thus discover equal numbers of objects for $\alpha\simeq0.5$,
and the deep search method discovers more objects for all steeper 
slopes.
The point here is that the pencil-beam method uses {\it all} the
available photons to contribute to the depth of the survey, whereas
the classical method uses only one-third, since it is the 
exposure limit of a single image that determines 
the depth of the classical survey.
For $\alpha\simeq0.7$ our pencil-beam method discovers
more than twice as many objects per night.
Of course, most of these objects are near the magnitude limit of the survey
and are thus not easily recoverable in order to monitor and improve
their orbits.  
The deep method is thus better suited to study the large-scale 
structure of the belt; for example, by compiling better statistics
on the number of objects as a function of ecliptic latitude.

Our best estimate for the luminosity function implies a surface density 
at magnitude $R\simeq29$ (radius $\sim$10 km at 45 AU) 
of $\sim4\times10^4$ TNOs per square degree which, assuming
a belt of latitudinal extent $\pm 15^\circ$ implies $4\times10^9$
Kuiper Belt objects from 30 to 50 AU, in rough agreement with previous
estimates based on the number of short-period comets (Levison and 
Duncan 1997). 
However, the uncertainty of the radii of a typical short-period
comets and the steepness of the luminosity function results in it
being very easy to tune this number by small variations of the
magnitude of a `comet' in the Kuiper Belt.
Nevertheless, this steeper luminosity function implies that
the Kuiper Belt could be the current source of the Jupiter-family 
comets, although a component of objects coming from the `scattered 
disk' cannot be ruled out (see Duncan and Levison 1997).

The lack of detections of Centaurs or objects beyond 50 AU in
our Palomar data is consistent with a simple extrapolation
of the size and radial distributions.
A doubling or tripling of the number of TNOs needs to occur without
discoveries outside 50 AU before there is a convincing problem with 
the lack of detections here.
A single large (30\arcmin$\times$30\arcmin) pencil-beam survey in
the ecliptic to $R>26$ should answer the question.


\acknowledgments

{\bf Acknowledgments:}
We thank Martin Duncan, Hal Levison, and Scott Tremaine for useful 
discussions, and E. Fletcher and A. Fitzsimmons for cooperation
in the discovery of 1997 GA45.  JJK is grateful for the financial support
of the Fund for Astrophysical Research.
We reserve special thanks for Brian Marsden, who provided orbit 
computations for candidates with a smile and at low cost. 



\end{document}